\documentclass[12pt]{article}

\usepackage{sw20jdoc}
\usepackage{fullpage}
\usepackage{amsmath}
\usepackage{amsfonts}
\usepackage{amssymb}
\usepackage{graphicx}
\usepackage{color}
\usepackage{lineno}

\providecommand{\U}[1]{\protect\rule{.1in}{.1in}}
\renewcommand{\baselinestretch}{1.5}
\newpagefalse

\begin{document}
\Title Representation of equatorial waves by a single scalar field and the effect of the horizontal component of the Earth's rotation

\Author Dennis Wilson Moore\footnote{Affiliated Professor of Oceanography, University of Washington, Seattle, USA; dwmoore@uw.edu}

\Date

\begin{abstract}%

\section{Abstract}

A new representation for equatorial waves in terms of a single scalar field is used to demonstrate the influence of the (usually neglected) horizontal component of the earth's rotation vector $\boldsymbol{\vec\Omega}$ on the dispersion of the waves. It is shown that, in the case of constant Brunt-V\"ais\"al\"a frequency $N$, the dispersion relation is unaltered when the horizontal component of $\boldsymbol{\vec\Omega}$ is included.
\bigskip

\noindent\textbf{Keywords:} equatorial waves, ocean physics, wave dispersion%

\end{abstract}%

\section{Introduction}
\label{sec:intro}

\textcolor{black}{Needler and LeBlond (1973)} have investigated the influence of the horizontal component of the earth's rotation on long period waves. For a thin shell of stratified fluid they concluded that \textquotedblleft the inclusion of the horizontal component of the Earth's rotation is found to have no noticeable effect on the dispersion relation of long period waves; its only influence is the introduction of a vertical phase shift in the motions.\textquotedblright\ Their analysis excluded the equatorial regions, although they remarked that \textquotedblleft this is not an essential restriction.\textquotedblright\ \textcolor{black}{Grimshaw (1975)} and \textcolor{black}{Stern (1975)} looked at the same problem for inertia-gravity waves, but again neither dealt specifically with equatorial solutions.

The purpose of this note is to investigate the influence of the horizontal component of the earth's rotation vector $\boldsymbol{\vec\Omega}$ on equatorial waves. In order to do so, the horizontal component of $\boldsymbol{\vec \Omega}$ is
approximated by a constant $\Omega_{h}$, and the vertical component is approximated by $\beta y/2$, which is the usual equatorial $\beta$-plane approximation \textcolor{black}{(Moore and Philander, 1977)}. This is the equatorial version of what \textcolor{black}{Grimshaw (1975)} refers to as a \textquotedblleft rational $\beta$-plane approximation.\textquotedblright

We introduce a new representation for equatorial waves in terms of a single scalar field, which is essentially a generalized streamfunction or potential function. The same sort of representation has been discovered independently by
\textcolor{black}{Pedro Ripa (1994)}. This representation may turn out to have a wide variety of applications to problems of rotating, stratified flow on a $\beta$-plane. The reason is its ability to express all velocity components in terms of a single scalar, which allows the boundary conditions to be written concisely, even for relatively complicated basin geometries.

This paper was originally written in 1993 and submitted to an editor. He rejected it without having it reviewed. I gave copies to a few friends, and was recently asked and encouraged to attempt to publish it. Special thanks to Dailin Wang for initiating this effort multiple times, and never giving up.

\section{Classical equatorial waves}
\label{sec:classical}

Consider a stably stratified ocean initially at rest. Let the background density distribution be $\rho_{B}=\rho_{0}+\bar{\rho}(z)$, and
\begin{equation}
N^{2}(z)=-\frac{g}{\rho_{0}}\frac{d\bar{\rho}}{dz}
\label{eq:N2}
\end{equation}
be the square of the Brunt-V\"ais\"al\"a frequency $N(z)$. First assume the Boussinesq and hydrostatic approximations, and neglect mixing and friction. The traditional equations for linear waves on an equatorial $\beta$-plane are
then
\begin{eqngroup}
\begin{equation}
u_{t}-\beta yv+\frac{1}{\rho_{0}}p_{x}=0,
\label{eq:u(trad)}
\end{equation}
\begin{equation}
v_{t}+\beta yu+\frac{1}{\rho_{0}}p_{y}=0,
\label{eq:v(trad)}
\end{equation}
\begin{equation}
u_{x}+v_{y}+w_{z}=0,
\label{eq:cont(trad)}
\end{equation}
\begin{equation}
p_{z}+\rho g=0,
\label{eq:w(trad)}
\end{equation}
and
\begin{equation}
\rho_{t}+w\bar{\rho}_{z}=0.
\label{eq:rho(trad)}
\end{equation}
\end{eqngroup}%
See, for example, \textcolor{black}{Moore and Philander (1977)}. Subscripts $x$, $y$, $z$ and $t$ in these, and subsequent equations, denote partial derivatives. The perturbation fields are the pressure $p$, the density $\rho$, the zonal velocity $u$, the meridional velocity $v$, and the vertical velocity $w$. The Coriolis parameter $f=\beta y$ is twice the locally vertical component of the earth's rotation vector $\boldsymbol{\vec\Omega}$, where the sine of the latitude
has been approximated by $y/R$, with $R$ denoting the radius of the earth, and $y$ measuring distance northward from the equator.

Equations (\ref{eq:cont(trad)})--(\ref{eq:rho(trad)}) may be combined, eliminating $\rho$ and $w$ in favor of $p$, to obtain
\begin{equation}
u_{x}+v_{y}=\frac{\partial}{\partial z}\left(\frac{p_{zt}}{\rho_{0}N^{2}
(z)}\right).
\label{eq:ux+vy(trad)}
\end{equation}

Eliminating $u$ between equations (\ref{eq:u(trad)}) and (\ref{eq:ux+vy(trad)}) results in the following equation involving $p$ and $v$:
\begin{equation}
\frac{1}{\rho_{0}}p_{xx}+\frac{1}{\rho_{0}}\frac{\partial}{\partial z}\left(\frac{p_{ztt}}{N^{2}(z)}\right) 
= \beta yv_{x}+v_{yt}.
\label{eq:pxx(trad)}
\end{equation}

Now suppose $v\not\equiv 0$, and assume we can find a scalar field $\Psi$ such that%
\begin{equation}
v=\Psi_{xx}+\frac{\partial}{\partial z}\left(  \frac{\Psi_{ztt}}{N^{2}(z)}\right).
\label{eq:vsol(trad)}
\end{equation}

Then from equation (\ref{eq:pxx(trad)}) we obtain
\begin{equation}
\frac{p}{\rho_{0}}=\beta y\Psi_{x}+\Psi_{yt},
\label{eq:psol(trad)}
\end{equation}
and by substituting into equation (\ref{eq:u(trad)}) we find
\begin{equation}
u=-\Psi_{xy}+\beta y\frac{\partial}{\partial z}\left(\frac{\Psi_{zt}}{N^{2}\left(  z\right)}\right) .
\label{eq:usol(trad)}
\end{equation}

Equations (\ref{eq:vsol(trad)}), (\ref{eq:psol(trad)}) and (\ref{eq:usol(trad)}) provide expressions for $v$, $p$, and $u$ in terms of a single scalar field $\Psi$. Equations (\ref{eq:w(trad)}) and (\ref{eq:rho(trad)}) may be used to write $w$ and
$\rho$ in terms of $\Psi$. For example,
\begin{equation}
w=-\frac{\beta y\Psi_{xzt}+\Psi_{yztt}}{N^{2}(z)}.
\label{eq:wsol(trad)}
\end{equation}

So far, we have not used the meridional velocity equation (\ref{eq:v(trad)}) at all! Substituting from equations (\ref{eq:vsol(trad)}), (\ref{eq:psol(trad)}), and (\ref{eq:usol(trad)}) into equation (\ref{eq:v(trad)}) gives the governing equation for $\Psi$:%
\begin{equation}
\Psi_{xxt}+\Psi_{yyt}+\beta\Psi_{x}+
\left( \frac{\partial^{2}}{\partial t^{2}}+\beta^{2}y^{2}\right)  
\left[ \frac{\partial^{2}}{\partial z\partial t}\left( \frac{1}{N^{2}(z)}\Psi_{z}\right) \right] = 0.
\label{eq:Psi(trad)}
\end{equation}
Equation (\ref{eq:Psi(trad)}) is the same equation that governs the meridional velocity $v$ in the usual equatorial wave theory.

If we are studying a single moving layer, or one baroclinic mode of the continuously stratified system, the above development works equally well. We simply replace $(N^{-2}\Psi_{z})_{z}$ with $-c^{-2}\Psi$, where $c$ is the Kelvin wave speed for the layer in question. \textcolor{black}{Pedro Ripa (1994)} has independently discovered the same scalar representation for this single layer case.

Since $u$, $v$, and $w$ are the components of a non-divergent vector, they may be written as the components of the curl of another vector $\boldsymbol{\vec A}$. If the components of $\boldsymbol{\vec A}$ are $A$, $B$, and $C$, then we find from
equations (\ref{eq:vsol(trad)})--(\ref{eq:usol(trad)}) that $A=N^{-2}\Psi_{ztt}$, $B=-\beta yN^{-2}\Psi_{zt}$, and $C=-\Psi_{x}$. The gradient of any scalar field may be added to $\boldsymbol{\vec A}$ without changing the velocity.

One question to investigate immediately is whether there exist non-trivial solutions for $\Psi$ for which the corresponding $u$, $v$, and $p$, derived from eqations (\ref{eq:vsol(trad)})--(\ref{eq:usol(trad)}), are identically zero. It is easy to see that such \textquotedblleft null\textquotedblright\ solutions exist. Assume a $\Psi$ field of the form%
\begin{equation}
\Psi = \exp[i(kx-\omega t)]\Upsilon(y,z),
\label{eq:Psi(periodic,trad)}
\end{equation}
and substitute (\ref{eq:Psi(periodic,trad)}) into (\ref{eq:psol(trad)}) with $p=0$. Then%
\begin{equation}
\Upsilon_{y} = \frac{k}{\omega}\beta y\Upsilon,
\label{eq:Upsilon(trad)}
\end{equation}
which means
\begin{equation}
\Psi = \exp\left[ i(kx-wt)+\frac{\beta ky^{2}}{2\omega}\right] \phi(z).
\label{eq:Psi(periodic,trad)2}
\end{equation}
Equation (\ref{eq:vsol(trad)}) with $v=0$, or equation (\ref{eq:usol(trad)}) with $u=0$, then gives
\begin{equation}
\frac{k^2}{\omega^2}\phi + \frac{d}{dz}\left( \frac{\phi_{z}}{N^{2}(z)} \right) = 0
\label{eq:phi(z)(trad)}
\end{equation}
as the equation for the vertical structure $\phi(z)$. Note from equation (\ref{eq:Psi(periodic,trad)2}) that if the phase velocity $k/\omega$ is positive (\textit{i.e.}, eastward), $\Psi$ grows exponentially away from the equator. If $k/\omega$ is negative (\textit{i.e.}, westward), $\Psi$ is Gaussian in $y$. So, we find that there are \textquotedblleft null\textquotedblright\ solutions for $\Psi$, which have the structure of a Kelvin wave going the wrong direction. That is, the eastward-propagating solutions are unbounded at high latitude and the westward-propagating ones are equatorially trapped.

It is easy to see that the separable (non-null) solutions for $\Psi$ give the known results for equatorially trapped baroclinic waves. In particular, in a constant $N$ ocean, a vertically standing or propagating solution with vertical wave number $\lambda$ has $v = \left( \lambda^2\omega^2/N^2-k^2\right) \Psi$. Therefore, all the equatorial wave solutions with $v\not = 0$ and $kN\not = \pm\lambda\omega$\thinspace can be represented in this way. Since $v$ and $\Psi$ both obey equation (\ref{eq:Psi(trad)}), any $v$ field will serve as a $\Psi$ field.

\section{Vertical acceleration, friction, and the horizontal component $\Omega_h$}
\label{sec:general}

Now consider the following generalization for the case of constant $N$. We want to include the effects of friction, vertical mixing of density, vertical accelerations (non-hydrostatic effects), and the locally horizontal component of the earth's rotation $\Omega_{h}$. This discussion is limited to constant $N$.

Let
\begin{equation}
D_{u} = D_{v} = D_{w} = \frac{\partial}{\partial t}
-\nu\frac{\partial^{2}}{\partial z^2}
\label{eq:DuDvDw}
\end{equation}
be the vertical diffusion operator for momentum, and
\begin{equation}
D_{\rho} = \frac{\partial}{\partial t}-\kappa\frac{\partial^2}{\partial z^2}
\label{eq:Drho}
\end{equation}
be the corresponding operator for diffusion of density. We assume the diffusion coefficients $\nu$ and $\kappa$ are both constant. The momentum diffusion operators $D_{u}$, $D_{v}$, and $D_{w}$ are all the same, but are labelled differently according to the momentum equation in which they appear. Thus, for example, the effect of the hydrostatic approximation may be seen be setting $D_{w}=0$ but retaining $D_{u}$ and $D_{v}$.

Let $\gamma=2\Omega_h$ be the Coriolis parameter for the locally horizontal component of earth's rotation at the equator. We begin with the following equations \textcolor{black}{(Moore and Philander, 1977)}:
\begin{eqngroup}
\begin{equation}
D_{u}u + \gamma w - \beta yv+\frac{1}{\rho_{0}}p_{x} = 0,
\label{eq:u(gen)}
\end{equation}
\begin{equation}
D_{v}v + \beta yu + \frac{1}{\rho_{0}}p_{y} = 0,
\label{eq:v(gen)}
\end{equation}
\begin{equation}
D_{w}w - \gamma u + \frac{1}{\rho_{0}}p_{z} + \frac{1}{\rho_{0}}\rho g = 0,
\label{eq:w(gen)}
\end{equation}
\begin{equation}
u_{x}+v_{y}+w_{z}=0,
\label{eq:cont(gen)}
\end{equation}
and
\begin{equation}
D_{\rho}\rho + w\bar{\rho}_{z}=0.
\label{eq:rho(gen)}
\end{equation}
\end{eqngroup}
We may eliminate $\rho$ between equations (\ref{eq:w(gen)}) and (\ref{eq:rho(gen)}) to obtain%
\begin{equation}
\left( D_{\rho}D_{w}+N^{2}\right)w - \gamma D_{\rho}u + 
D_{\rho}\frac{\partial}{\partial z}\left(  \frac{p}{\rho_0}\right) = 0.
\label{eq:wup(gen)}
\end{equation}

Working with equations (\ref{eq:u(gen)}), (\ref{eq:cont(gen)}) and (\ref{eq:wup(gen)}) in a manner completely analogous to what we did in Section \ref{sec:classical}, we find the following representation for $v$, $u$, $w$, $\rho$ and $p$ in terms of a scalar field $\Phi(x,y,z,t)$:
\begin{eqngroup}
\begin{equation}
v = D_{\rho}D_{u}\Phi_{zz} + \left( D_{\rho}D_{w}+N^{2}\right) \Phi_{xx},
\label{eq:vsol(gen)}
\end{equation}
\begin{equation}
u = D_{\rho}\beta y\Phi_{zz}+D_{\rho}\gamma\Phi_{yz}-(D_{\rho}D_{w}+N^{2})\Phi_{xy},
\label{eq:usol(gen)}
\end{equation}
\begin{equation}
w = -D_{\rho}(\beta y\Phi_{xz}+\gamma\Phi_{xy}+D_{u}\Phi_{yz}),
\label{eq:wsol(gen)}
\end{equation}
\begin{equation}
\rho = \bar{\rho}_{z}(\beta y\Phi_{xz}+\gamma\Phi_{xy}+D_{u}\Phi_{yz}),
\label{eq:rhosol(gen)}
\end{equation}
and
\begin{equation}
\frac{p}{\rho_{0}} = \left( D_{\rho}D_{w}+N^{2}\right)
(\beta y\Phi_{x}+D_{u}\Phi_{y})+\gamma D_{\rho}(\beta y\Phi_{z}+\gamma\Phi_{y}).
\label{eq:psol(gen)}%
\end{equation}%
\end{eqngroup}%

The $\Phi$ field in these equations has dimensions $L^{3}T$, which is $T^{2}$ times the dimensions of $\Psi$ in the previous section. Since $N^{2}$ is constant, $\Psi$ and $\Phi$ are simply proportional, that is, $\Psi=N^{2}\Phi$. The solenoidal velocity field is now given as the curl of a vector field $\boldsymbol{\vec A}$ with components $A=D_{\rho}D_{u}\Phi_{z}$, $B=-D_{\rho}(f\Phi_{z}+\gamma\Phi_{y})$, and $C=-\left(D_{\rho}D_{w}+N^{2}\right)\Phi_{x} $, where $f=\beta y$. The easiest way to see that equations (\ref{eq:vsol(gen)})--(\ref{eq:psol(gen)}) satisfy (\ref{eq:u(gen)}) and (\ref{eq:w(gen)})--(\ref{eq:rho(gen)}) is by direct substitution. Substitution into the meridional velocity equation (\ref{eq:v(gen)}) gives a single equation for $\Phi$, with $f=\beta y$:
\begin{equation}
(D_{\rho}D_{w}+N^{2})\left( D_{v}\Phi_{xx}+D_{u}\Phi_{yy}+\beta\Phi_{x}\right) 
+ D_{v}D_{u}D_{\rho}\Phi_{zz} 
+ D_{\rho}\left( f\frac{\partial}{\partial z}+\gamma\frac{\partial}{\partial y}\right)^{2}\Phi = 0.
\label{eq:Phi(gen)}
\end{equation}
This is the generalization for constant $N$ of equation (\ref{eq:Psi(trad)}), to include vertical friction ($\nu$), vertical mixing of density ($\kappa$), vertical acceleration, and the horizontal Coriolis term ($\gamma$).

In this exposition, the quantities $\beta$, $\gamma$, $N$, $\nu$, and $\kappa$ have all been assumed constant, but no further approximation has been made in going from equations (\ref{eq:u(gen)})--(\ref{eq:rho(gen)}) to equation (\ref{eq:Phi(gen)}). This means that the representation in terms of $\Phi$ derived here is applicable to a wide variety of problems. One example follows.

\section{The effects of vertical acceleration and $\Omega_{h}$ on equatorial waves}
\label{sec:eqwaves}

Let us investigate the possible equatorial waves of this system by assuming plane wave solutions for $\Phi$ of the form
\begin{equation}
\Phi = \exp\left[ i(kx+\lambda z-\omega t)\right]\phi(y).
\label{eq:Phi(periodic)}
\end{equation}
We seek inviscid non-diffusive solutions, so we take $\nu=\kappa=0$. Then
\begin{equation}
D_{u} = D_{v} = D_{w} = D_{\rho} = \frac{\partial}{\partial t}\rightarrow-i\omega,
\label{eq:DuDvDw(inviscid)}
\end{equation}
and substitution of equation (\ref{eq:Phi(periodic)}) into equation (\ref{eq:Phi(gen)}) gives the $\phi$ equation:
\begin{equation}
\left(N^{2}-\omega^{2}\right)\phi_{yy} 
+ \left( i\lambda\beta y+\gamma\frac{d}{dy}\right)^{2}\phi
= \left[ \left(  N^{2}-\omega^{2}\right)\left( k^{2}+\beta\frac{k}{\omega}\right)
         - \omega^{2}\lambda^{2}\right]\phi.
\label{eq:phi(y)}
\end{equation}
By analogy with classical equatorial waves we seek solutions to equation (\ref{eq:phi(y)}) of the form
\begin{equation}
\phi = \exp\left(-\frac{ay^{2}}{2}\right)P(y)
\label{eq:phi(y)sol}
\end{equation}
where $P$ is a polynomial in $y$ and the real part of $a$ is positive to assure the solutions are equatorially trapped. Substituting equation (\ref{eq:phi(y)sol}) into equation (\ref{eq:phi(y)}) and equating the coefficient of $y^{2}P$ to zero gives an equation for $a$,
\begin{equation}
a^{2}\left(N^{2}-\omega^{2}\right) + \left(i\lambda\beta-\gamma a\right)^{2}=0.
\label{eq:a}
\end{equation}
The solution with $Re(a)>0$ is
\begin{equation}
a = \frac{\beta\left\vert\lambda\right\vert \sqrt{N^{2}-\omega^{2}}+ i\beta\gamma\lambda}
         {N^{2}+\gamma^{2}-\omega^{2}},
\label{eq:asol}
\end{equation}
where we have assumed $\left\vert\omega\right\vert<N$. If we define a dimensional latitudinal variable $\eta$ by
\begin{equation}
\eta^{2} = \frac{\beta\left\vert\lambda\right\vert \sqrt{N^{2}-\omega^{2}}}
                {N^{2}+\gamma^{2}-\omega^{2}}y^{2},
\label{eq:eta2}
\end{equation}
and regard $P$ as a function of $\eta$, the resulting equation for $P$ is
\begin{equation}
\frac{d^{2}P}{d\eta^{2}}-\left(2\eta\frac{d}{d\eta}+1\right)P 
= \Lambda P
= \frac{\left(N^{2}-\omega^{2}\right)\left(k^{2}+\beta k/\omega\right)-\omega^{2}\lambda^{2}}
       {\beta\left\vert \lambda\right\vert \sqrt{N^{2}-\omega^{2}}} P.
\label{eq:P}%
\end{equation}
This is the Hermite equation, see, for example, \textcolor{black}{Wiener (1933), equation (6.02)}. The polynomial solution with leading term $\eta^{n}$ has eigenvalue $\Lambda=-(2n+1)$, which gives the dispersion relation%
\begin{equation}
\left(N^{2}-\omega^{2}\right)\left(k^{2}+\frac{\beta k}{\omega}\right)
+(2n+1)\beta\left\vert \lambda\right\vert \sqrt{N^{2}-\omega^{2}}
=\omega^{2}\lambda^{2}.
\label{eq:disprel}%
\end{equation}
The effect of the vertical acceleration is to replace $N^{2}$ in the hydrostatic case by $N^{2}-\omega^{2}$ in the non-hydrostatic case, a well known result from internal wave theory. The locally horizontal component of the earth's rotation vector has no effect on the dispersion relation for the equatorial waves, since $\gamma$ does not appear in the equation (\ref{eq:disprel}). The only effect of the terms involving $\gamma$ is to slightly broaden the scale of the meridional fields, and to introduce an imaginary component in the Gaussian envelope. That is, the quantity $a$ in
equation (\ref{eq:phi(y)sol}) is complex, with an imaginary part proportional to $\gamma$, as shown by the solution (\ref{eq:asol}). Constant phase surfaces for $\Phi$ are given by
\begin{equation}
kx+\lambda\left(  z-\frac{\gamma\beta y^{2}/2}{N^{2}+\gamma^{2}-\omega^{2}}\right) = const.
\label{eq:phaselines}%
\end{equation}
The bending of phase lines on surfaces of constant $z$, depending on the sign of $\lambda$, destroys the vertical separability of the problem for an ocean bounded between $z=0$ and $z=-H$. That is, we cannot satisfy $w=0$ by superposing upward- and downward-propagating solutions with the same $\left\vert\lambda\right\vert$.

\section{Equatorial Kelvin waves}
\label{sec:kelvinwaves}

We have assumed $v\not=0$ in most of our discussion. There is an important class of equatorial waves for which $u$ and $p$ are non-zero but $v\equiv0$, namely, the equatorial Kelvin waves. Here, we simply write down the standard Kelvin wave results, show what $\Omega_{h}$ does to them, and demonstrate that they also can be written in terms of the scalar field. 

The basic Kelvin wave solution for equations (\ref{eq:u(trad)})--(\ref{eq:rho(trad)}) has
\begin{equation}
u = \frac{p}{c\rho_{0}} = U_{0}\exp\left[ik(x-ct)-\frac{\beta y^{2}}{2c}\right]\phi(z),
\label{eq:uKW}%
\end{equation}
where
\begin{equation}
\frac{1}{c^{2}}\phi + \frac{d}{dz}\left(\frac{\phi_{z}}{N^{2}(z)}\right)=0
\label{eq:phi(z)KW}%
\end{equation}
describes the vertical structure of the Kelvin waves. The waves propagate eastward with phase speed $k/\omega = c$. For the case $N=const$, the Kelvin wave may be written as%
\begin{equation}
u = U_{0}\exp\left[i(kx+\lambda z-\omega t)-\frac{\beta\left\vert
\lambda\right\vert y^{2}}{2N}\right]  ,
\label{eq:uKW2}%
\end{equation}
with $\omega/k=N/\left\vert \lambda\right\vert $. With the vertical acceleration and $\gamma$ terms retained, the Kelvin wave solution becomes
\begin{equation}
u = U_{0}\exp\left[i(kx+\lambda z-\omega t)-\frac{\beta y^{2}}{2}
\frac{\sqrt{N^{2}-\omega^{2}}\left\vert \lambda\right\vert +i\gamma\lambda}{N^{2}+\lambda^{2}-\omega^{2}}\right],\label{eq:uKW3}
\end{equation}
with%
\begin{equation}
\frac{\omega}{k}=\frac{\sqrt{N^{2}-\omega^{2}}}{\left\vert \lambda\right\vert}.
\label{eq:KWdisprel}
\end{equation}
So, again we see that $\gamma$ has no effect on the dispersion relation and simply adds a phase shift to the wave structure. The vertical acceleration has the usual effect of replacing $N^{2}$ by $N^{2}-\omega^{2}$.

The Kelvin wave solution in equation (\ref{eq:uKW}) is in fact derivable from a $\Psi$ field of the form
\begin{equation}
\Psi=\Psi_{0}\phi(z)\exp\left[  ik(x-ct)+\frac{\beta y^{2}}{2c}\right]
\int_{0}^{y}\exp\left(  -\beta\zeta^{2}/c\right)  d\zeta.\label{Psi(K)sol}
\end{equation}
Although this $\Psi$ field is unbounded at high latitudes, the velocity field derived from it is equatorially trapped, and is an acceptable equatorial solution to equations (\ref{eq:u(trad)})--(\ref{eq:rho(trad)}).

\section{Conclusions}

The effect of the horizontal component of the earth's rotation on equatorial waves has been investigated. There is no effect on the dispersion relation for the waves, but there is an effect on the spatial structure of the wave field. The usual Gaussian envelope for the wave fields now has an imaginary part in the exponent, so that constant phase surfaces are curved in the $yz$-plane. The imposition of the boundary condition $w = 0$ on $z = const$ makes the problem non-separable. These are basically the same conclusions that \textcolor{black}{Needler and LeBlond (1973)} found for long period waves away from the equator.

The analysis is carried out by using a single scalar field $\Psi$ to represent the velocity components, pressure and density perturbations. This scalar representation appears to be applicable to a wide variety of problems.

\acknowledge

\section{Acknowledgments}

This work was supported under a National Science Foundation grant (OCE9019580) to the University of Hawai`i. Lew Rothstein, Eric Firing and Jay McCreary contributed many useful discussions. Gary Mitchum, Steve Chiswell, Jim O'Brien, and a host of their coevals once tried to get me to find Kelvin waves going the \textquotedblleft wrong way,\textquotedblright\ as the \textquotedblleft null\textquotedblright\ solution for the scalar $\Psi$ does in this problem. Pedro Ripa's independent discovery of the scalar representation provided the impetus to write up this work. Dailin Wang, Jay McCreary, Ted Durland and Hristina Hristova were extremely helpful with the preparation of this manuscript.




\end{document}